\documentclass[aps,prl,showpacs,twocolumn,superscriptaddress]{revtex4-1}

\usepackage{amsmath}    
\usepackage{graphicx}   
\usepackage{verbatim}   
\usepackage{color}      
\usepackage{subfigure}  
\usepackage{hyperref}   
\definecolor{orange}{rgb}{1,0.5,0}

\begin{document}

\title{A quantum parametric oscillator with trapped ions}
\author{Shiqian Ding}
\author{Gleb Maslennikov} 
\author{Roland Habl{\"u}tzel}
\affiliation{Centre for Quantum Technologies, National University of Singapore, 3 Science Dr 2, 117543, Singapore}
\author{Huanqian Loh}
\affiliation{Centre for Quantum Technologies, National University of Singapore, 3 Science Dr 2, 117543, Singapore}
\author{Dzmitry Matsukevich}
\affiliation{Centre for Quantum Technologies, National University of Singapore, 3 Science Dr 2, 117543, Singapore}
\affiliation{Department of Physics, National University of Singapore, 2 Science Dr 3, 117551, Singapore}

\date{\today}

\begin{abstract}
A system of harmonic oscillators coupled via nonlinear interaction 
is a fundamental model in many branches of physics, from biophysics to electronics and
condensed matter physics. In quantum optics, weak nonlinear interaction between light modes 
has enabled, for example, the preparation of squeezed states of light 
and generation of entangled photon pairs~\cite{Kwiat_1995,kimble_squeezed_1986}.
While strong nonlinear interaction between the modes has been realized in circuit QED systems~\cite{lecocq_2012,holland_crosskerr_2015}, 
achieving significant interaction strength on the level of single quanta in other physical systems 
remains a challenge~\cite{feizpour_crosskerr_2015,langford_efficient_2011}.
Here we experimentally demonstrate such interaction that is equivalent 
to photon up- and down-conversion using normal modes of 
motion in a system of two Yb ions~\cite{james_complete_2003,Nie_2009}. The nonlinearity is
induced by the intrinsic anharmonicity of the Coulomb interaction between the ions 
and can be used to simulate fully quantum operation of 
a degenerate optical parametric oscillator~\cite{Mandell_book}. 
We exploit this interaction to directly measure the parity and Wigner functions of ion motional states. The nonlinear coupling, 
combined with near perfect control of internal and motional states of trapped ions, 
can be applied to quantum computing~\cite{langford_efficient_2011,andersen_hybrid_2015}, 
quantum thermodynamics~\cite{quantum_enhanced_fridge_2014,schmidt_kaler_engine_2014}, 
and even shed some light on the quantum information aspects of Hawking radiation~\cite{2010_hawking_trilinear}.
\end{abstract}

\maketitle

Ions in a Paul trap experience a pseudopotential that is harmonic to a high degree and their motion is usually approximated by a set of normal modes that do not interact with each other. Coulomb interaction between the ions 
is, however, nonlinear and can introduce coupling between the modes 
and anharmonicity to the ion motion. The linear coupling of motional 
modes due to mutual Coulomb repulsion of ions was previously demonstrated 
in quantum regime~\cite{wineland_oscillator,blatt_oscillator}, 
where the ions were trapped in independent 
potential wells. The higher order terms in the Coulomb interaction lead, 
for example, to cross Kerr-type nonlinear coupling that results in a shift of the normal
mode frequencies~\cite{james_complete_2003,Nie_2009} which has been experimentally
observed~\cite{Roos_shift_2008}.

In this Letter, we engineer the nonlinear interaction between  
modes of motion similar to a degenerate parametric oscillator at the single-phonon level. 
We consider a system of two ions with the same mass $m$ and charge $e$ in a 
linear Paul trap that is characterized by the single-ion secular frequencies $\omega_x$, $\omega_y$, $\omega_z$ ~\cite{james_complete_2003,Nie_2009}. 
The potential energy of the system has the form 
\begin{eqnarray} \label{eq:energy}
V &=& m \omega_x^2 (X^2+x^2) + m \omega_y^2 (Y^2 +y^2)  \\ \nonumber
  & & + m \omega_z^2 (Z^2+ z^2) + \frac{e^2}{8 \pi \epsilon_0} \frac{1}{\sqrt{x^2 + y^2 + z^2}} \, ,
\end{eqnarray}
where $\epsilon_0$ is the permittivity of free space, $X, Y, Z$ are the center-of-mass
coordinates, and $x, y, z$ are half the separation between the ions along the direction
of principle trap axes.

When $\omega_z < (\omega_x, \omega_y$), the ions crystallize along the axial ($z$) direction at an equilibrium distance 
$z_0$  from the trap center.
According to Eq.~\ref{eq:energy}, the motion of the center-of-mass modes is harmonic, 
but the out-of-phase modes are coupled to each other by the 
Coulomb interaction. For small axial displacement $u=z-z_0$ and keeping only terms up to the 
third order that contribute to the coupling between the $x$ and $z$ modes, the potential energy becomes~\cite{Nie_2009}
\begin{equation}\label{eq:pot}
V = m\omega_{r}^2 x^2 + m\omega_s^2 u^2 +\frac{m\omega_s^2}{z_0} x^2 u + ... \, .
\end{equation}
Here $\omega_s=\sqrt{3}\,\omega_z,\,\omega_r=\sqrt{\omega_x^2-\omega_z^2}$ are the axial (``stretch") and radial (``rocking") 
mode frequencies for the out-of-phase motion. If the trap frequencies are chosen such that 
$\omega_s \simeq 2 \omega_r$, we can apply the standard transformations $\hat{x}=(\hbar/4m\omega_r)^{1/2}(\hat{a}+\hat{a}^\dagger)$, 
$\hat{u}=(\hbar/4m\omega_s)^{1/2}(\hat{c}+\hat{c}^\dagger)$ and express the Hamiltonian in the rotating wave approximation as 
\begin{equation}\label{eq:ham}
\hat{H}=\hbar\omega_r \hat{a}^{\dagger}\hat{a} + \hbar\omega_s \hat{c}^{\dagger}\hat{c}+\hbar\xi\left(\hat{a}^{\dagger\,2} \hat{c}+\hat{a}^2 \hat{c}^{\dagger}\right),
\end{equation}
where $\hat{c}^{\dagger},\hat{c},(\hat{a}^{\dagger},\hat{a})$ are the phonon creation and annihilation operators in axial (radial) mode. The first two terms in Eq.~\ref{eq:ham} describe harmonic motion
in the axial (radial) mode with the frequency $\omega_s\,(\omega_r)$ and the third term 
couples these modes with the coupling coefficient given by 
\begin{equation}\label{eq:coupling}
\xi= \frac{1}{8 z_0}\sqrt{\frac{\hbar \omega_s^3}{m\,\omega_r^2}} \, .
\end{equation}
The coupling is nonlinear: one phonon from the axial mode is converted 
into a pair of phonons in the radial mode and vice versa.

\begin{figure}[btp!]
\centering
\includegraphics[width=\columnwidth]{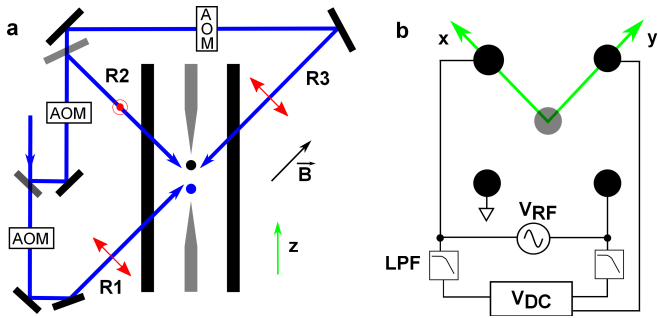}
\caption{\label{fig:setup} {\bf Experimental setup.} 
{\bf a.} Schematic of the experimental setup. 
The two-ion crystal is aligned along the axial 
direction of the trap. One of the Yb ions is in the $S_{1/2}$ ground state and interacts
with Doppler cooling and Raman beams. The other ion is prepared in a dark metastable
$^2F_{7/2}$ state and is sympathetically cooled.
Three Raman beams are directed onto the ions in such a way that the
motional states along the axial and radial directions are controlled 
by the R1-R2 pair and R2-R3 pair, respectively. The red labels show the polarizations of
Raman beams. The 7.0 Gauss magnetic field $\vec B$ is parallel to the R1 Raman beam.
{\bf b.} Trap electrode configuration. The trap consists of two end-cap electrodes 
separated by about 2.0\,mm and four rods. Each trap rod has a diameter of 0.5\,mm 
and the center-to-center separation is 0.9\,mm. 
The axial trapping frequency is controlled by the DC voltages applied to the end caps,
and the radial trapping potential is generated by 30\,MHz rf signal connected to the
diametrically opposite rods in the $x$ direction.  The fine control of radial motional
frequencies is achieved by adjusting offset voltages on the $x$
electrodes with two pairs of low pass filters (LPF) with different time constants (see 
Methods). Small additional DC voltages applied to the $x$ and $y$ electrodes
help to compensate stray electric fields. 
}
\end{figure} 

A schematic of the experimental setup is shown in Fig.~\ref{fig:setup}. 
Two $^{171}$Yb$^+$ ions are confined in a four-rod linear rf-Paul trap 
with the single-ion secular frequencies $(\omega_x, 
\omega_y, \omega_z)/2\pi=(0.99, 0.90, 0.75)$~MHz. 
To change the detuning $\delta=\omega_s - 2 \omega_r$, the axial 
trapping frequency remains fixed for all the experiments,
while the radial frequency is tuned by adjusting the DC voltages applied to the $x$
electrodes (see Fig.~\ref{fig:setup}b). One of the ions is optically pumped 
into a metastable $^2F_{7/2}$ state and does not interact with the laser beams 
during the experiment.
At the beginning of every experimental sequence, the detuning 
is set at $\delta/2\pi=35~\mathrm{kHz}$, which is much larger than $\xi/2\pi$,
effectively decoupling two modes.
We then initialize all the motional modes of the two-ion crystal 
in the ground state by Doppler cooling followed by 
sideband cooling~\cite{resolvedsbc}. The spin-motion coupling for the sideband 
cooling is achieved by driving frequency-comb-assisted Raman
transitions~\cite{Hayes_2010} (see Methods). The residual population $\bar{n}$ of all 
the motional modes after sideband cooling is well below 0.05 phonons. 

\begin{figure}[tbp!]
\centering
\includegraphics[width=\columnwidth]{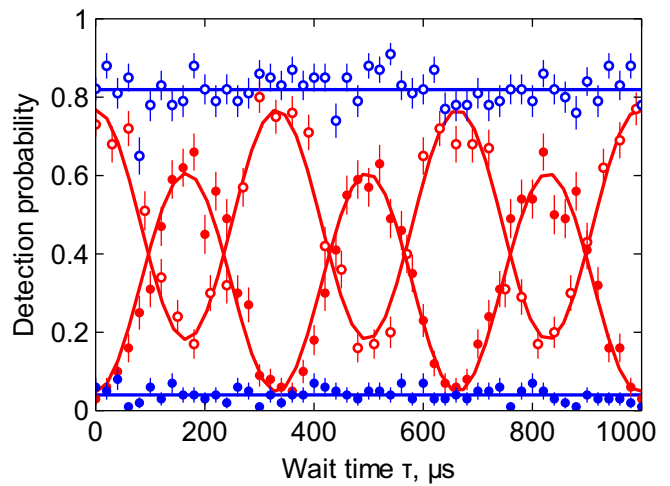}
\caption{\label{fig:onetotwo} 
{\bf Phonon state evolution in the axial (solid dots) and radial 
(open circles) modes}, where either one (blue) or two (red) 
phonons is initially added to the radial mode. The sinusoidal fit (red line) 
reveals the coupling rate to be 3.02$\pm 0.02$\,kHz. 
The straight fit lines show that there is 
negligible coupling between the modes when only one radial phonon is present. 
The error bars are statistical uncertainties that correspond to one 
standard deviation $\sigma$.
The oscillations amplitude after about 10~ms is a factor of 0.39(7) compared 
to the initial amplitude, limited by the coherence time of the phonons in the radial mode.}
\end{figure}

To verify the nonlinearity of the coupling at the single phonon level,
we initially populate the radial mode with either one or two 
phonons as described in Methods section. We then change DC voltages applied 
to the rods of the trap, with time constants of around 
$20~\mu\mathrm{s}$, to bring the detuning $\delta$ to zero.  
After time $\tau$, we bring the detuning back to its initial 
value and check for the presence of phonons in the axial or radial mode. 
The results are presented in Fig.~\ref{fig:onetotwo}. 

We observe energy oscillations between the axial and radial modes 
only when the radial mode is initially prepared in the two-phonon Fock state. 
The measured oscillation frequency $3.02\pm 0.02$ kHz is compatible with the frequency 
$2 \sqrt{2} \xi / 2 \pi = 2.96$ kHz predicted by Eq.~\ref{eq:coupling}. 
The reduced visibility of the oscillation together with the small deviation 
of the measured coupling strength from theory can be attributed to two sources: the non-perfect
mapping of the motional state to the internal state of the ion, and the 
deviation from the resonance condition, compatible with the observed frequency drifts. 

The observed mode coupling is analogous to the up- and down-conversion of photons in nonlinear crystals. 
However, in contrast to the optical case, where the number of pump photons 
required to produce one photon pair is usually large~\cite{langford_efficient_2011}, 
the coupling strength here is much higher and this effect can be readily observed 
even at the single quantum level.

\begin{figure}[tbp!]
\centering
\includegraphics[width=1\columnwidth]{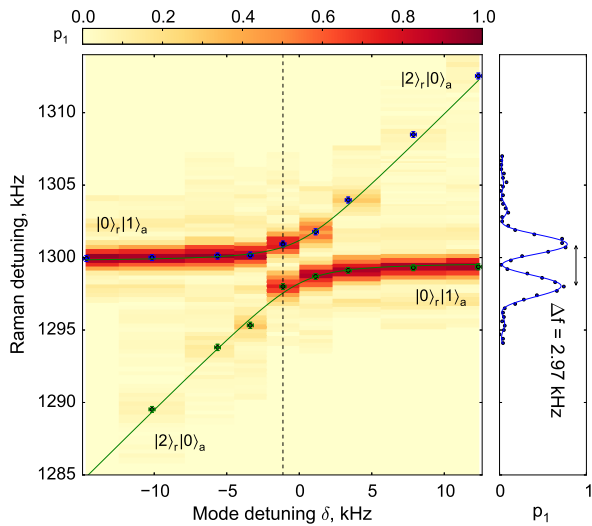}
\caption{\label{fig:anticrossing} 
{\bf Avoided crossing observed for the axial mode after sideband cooling.}
The left plot shows the probability $p_1$ of the transition from a ``dark"
$|F=0,m_F=0\rangle$ to ``bright" $|F=1,m_F=0\rangle$ state as a function of the detuning 
$\delta$ from the resonance condition, and the Raman detuning $\Delta$. 
We extract the coupling strength by measuring the mode splitting 
at resonance, as shown on the right panel. The splitting is measured 
to be $2.97 \pm 0.03 $\,kHz. Dots correspond to the measured frequencies 
at the peak centers and the solid lines show the eigenvalues 
of the Hamiltonian (\ref{eq:ham}). } 
\end{figure}

In order to further quantify this coupling, we probe 
the avoided crossing of the coupled modes of motion. 
Without the coupling ($\xi \rightarrow 0$), the bare energy eigenstates of the 
Hamiltonian in Eq.~\ref{eq:ham} are degenerate and cross when $\omega_s = 2 \omega_r$. 
The coupling mixes bare energy eigenstates of the system such that the new 
eigenstates have non-zero projection along both the axial and radial directions. 
This results in the mode splitting that we measure in the vicinity of blue sideband of axial mode. The results are shown in Fig.~\ref{fig:anticrossing}. 

If the detuning $\delta$ changes on a time scale much longer than 
the inverse coupling rate $2\pi/\xi$, the system remains in 
the same energy eigenstate, 
leading to the adiabatic evolution of motional states between
the radial and axial modes. In particular, the 
lowest energy eigenstate for $\omega_s > 2 \omega_r$,  i.e.\ $|n\rangle_r |0\rangle_a$,  
will evolve into the lowest energy eigenstate for $\omega_s < 2 \omega_r$. 
The latter eigenstate is $|0\rangle_r |n / 2 \rangle_a$ for even $n$  
and $|1\rangle_r |(n-1)/ 2 \rangle_a$ for odd $n$. 
Therefore, the adiabatic sweep enables direct parity measurement 
of the ion motional state: 
the absence or presence of a phonon in the radial mode after the sweep 
provides information about the parity of the initial state of this mode. 
We detect the phonon by mapping it onto the ion
internal state and then determine the expectation value of 
the parity operator  $\hat P | n \rangle_{r} = (-1)^n | n \rangle_{r}$ as
$\langle \hat P \rangle = (1 - 2 p_1 / \eta) $, 
where $\eta = 0.86$ is the phonon mapping efficiency (see Methods), and 
$p_1$ is the probability to find the ion in the ``bright" internal state after mapping.

The direct parity measurement opens the door to efficiently determining 
the Wigner function of a quantum state~\cite{Wigner_1932}. 
It was shown in ~\cite{Royer_1977,Englert_1993,Lutterbach_1997} that the Wigner function 
relates to the parity operator by
~\cite{Haroche_2002,Haroche_2008,Schoelkopf_2014,Leibfried_tomo_1996}
$$
W(\alpha) = \frac{2}{\pi} Tr[ D(-\alpha) \rho D(\alpha) \hat P] \, ,
$$
where $\rho$ is the density matrix and $D(\alpha)$ is the displacement operator. 
The state $ D(-\alpha) \rho D(\alpha) $ corresponds to a displacement of the 
state $\rho$ by the amount $-\alpha$ in phase space. 
The displacement can be carried out by applying a force to the 
ion with controlled phase and duration. 
After an adiabatic sweep of the radial trapping frequency with a time constant of 2~ms
(see Fig.~\ref{fig:setup}b), and mapping the radial phonon onto the internal state of the ion, the value of the Wigner 
function can be determined as $W(\alpha) = 2 \langle \hat P \rangle / \pi$. 

\begin{figure*}[tbp!]
\centering
\includegraphics[width=2.000 \columnwidth]{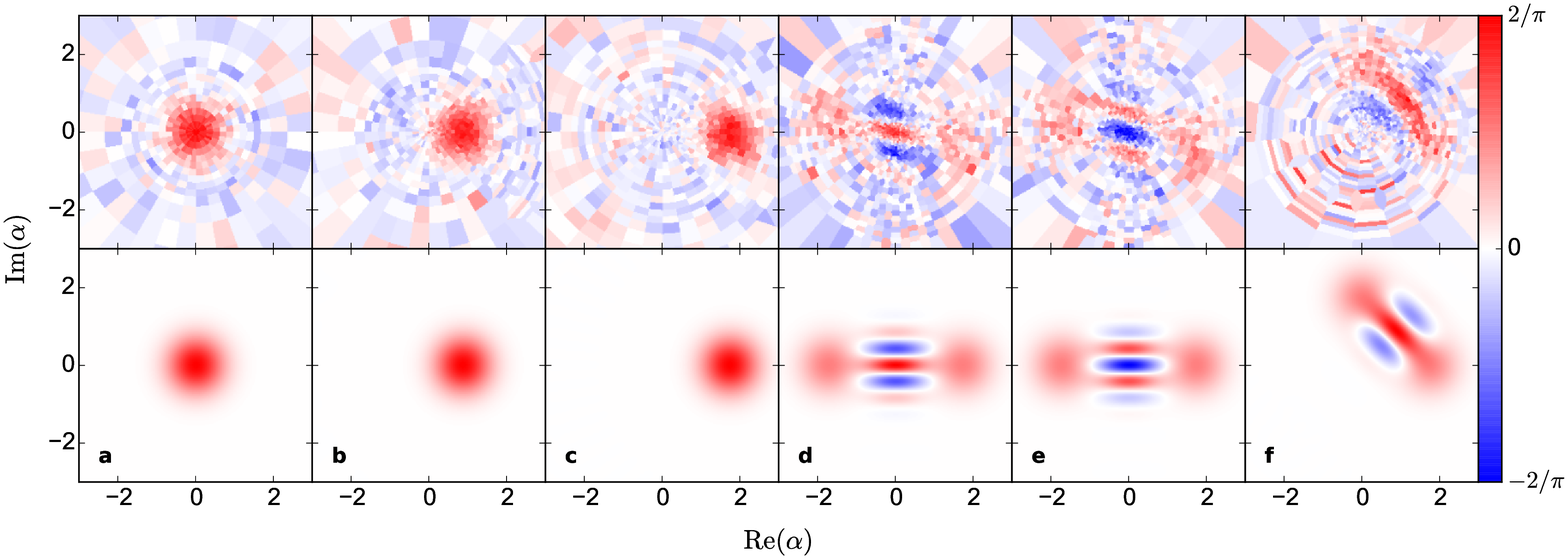}
\includegraphics[width=2.000 \columnwidth]{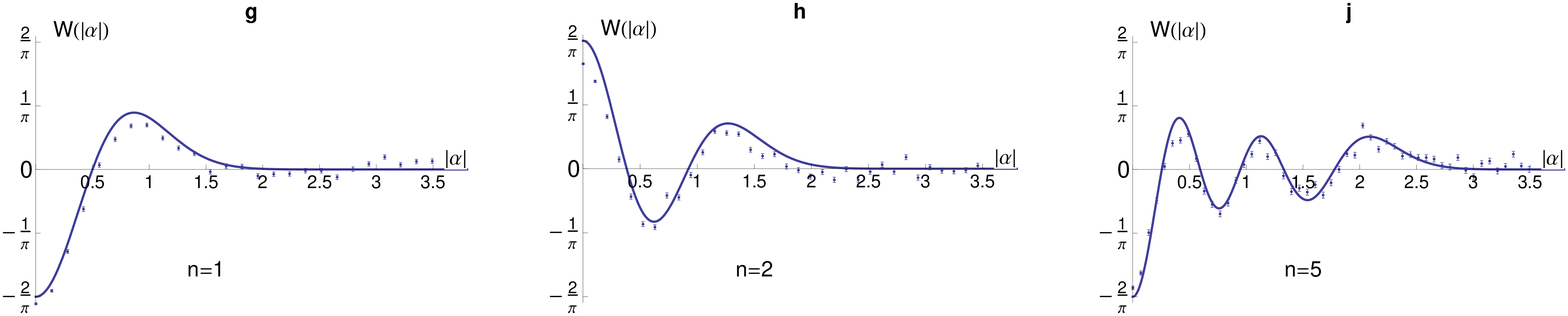}
\caption{\label{fig:wigner}{\bf Wigner functions for different quantum states.}
{\bf a-f.} Wigner functions (from left to right) for the vacuum state, coherent states 
$|\alpha\rangle_r$ for $\alpha$ = 0.87(3) and 1.73(6), Schr\"{o}dinger cat states 
$(|\alpha\rangle_r + |-\alpha\rangle_r)/\sqrt{2}$, 
$(|\alpha\rangle_r - |-\alpha\rangle_r)/\sqrt{2}$ and 
$(|\alpha\rangle_r - |i \alpha\rangle_r)/\sqrt{2}$
for $\alpha = 1.73(6)$.  
The top row corresponds to the experimental data, while the bottom row shows the calculated Wigner functions. 
{\bf g-j.} Wigner functions of the Fock states $|n\rangle_r$
averaged over the phase of $\alpha$ with $n=$1, 2, and 5. 
The solid lines show the theoretical prediction
 $W(|\alpha|) = 2 (-1)^n e^{-2 |\alpha |^2} L_n( 4 |\alpha|^2) / \pi$; 
 the points are experimentally measured values. 
 The error bars show $1\sigma$ statistical uncertainty. 
The negative measured values of the Wigner functions for the Schr{\"o}dinger cat and Fock 
states demonstrate the non-classical character of these states. }
\end{figure*} 

The measurement protocol of the Wigner function was tested on several quantum states,
prepared as described in the Methods section. 
Figures~\ref{fig:wigner}{a-f} show the Wigner functions for the vacuum state,
coherent states, and Schr\"{o}dinger cat states. The measured Wigner functions 
are shown in the top row and can be compared with the corresponding theoretical predictions plotted beneath them. Figures~\ref{fig:wigner}{g-j} show the Wigner functions 
for the Fock states with phonon number $n=1$, 2 and 5 as a function of $|\alpha|$. 
Some discrepancy between the theory and experiment is likely due to anharmonicity of
the motional mode that causes an amplitude-dependent rotation of 
the Wigner function about the origin.

In conclusion, we have demonstrated strong nonlinear coupling between the motional states of a two-ion crystal described by 
a Hamiltonian similar to degenerate optical parametric down-conversion. 
When combined with nearly deterministic phonon preparation and readout, 
both of which are readily available in an ion trap, the nonlinear coupling allows us to
directly measure the parity of the ion motional state and determine its Wigner 
function. This nonlinearity can be harnessed as a tool to implement hybrid quantum
computations that involve both discrete~\cite{langford_efficient_2011} and continuous
variables~\cite{andersen_hybrid_2015}, 
or to simulate a molecular BEC formation~\cite{jin_molecular_bec_2005}. 
One can also extend the this type of coupling to three modes in a three-ion crystal~\cite{james_complete_2003}, 
thereby realizing absorption refrigerators in the quantum 
regime~\cite{quantum_enhanced_fridge_2014}, 
or even simulating quantum information aspects of Hawking 
radiation~\cite{2010_hawking_trilinear}.

During preparation of this manuscript, we became aware of related work of Kienzler et al.~\cite{Kienzler_2015}, where a different method of the Wigner function reconstruction based on the  
measurements in a squeezed Fock basis was demonstrated.

We acknowledge discussions with Berthold-Georg Englert, Manas Mukherjee 
and Alex Kuzmich. This research was supported by the National Research Foundation and the
Ministry of Education of Singapore.

\section{Methods}
\subsection{Experimental setup}
The ions are Doppler cooled by a 369.53~nm laser of intensity 3~W/cm$^2$ and red-detuned
from the $^{2}S_{1/2},F=1 \rightarrow$ $^{2}P_{1/2},F=0$ transition.  
A 935.19~nm laser of intensity 25~W/cm$^2$ repumps the population from the 
long-lived  $^{2}D_{3/2}$ state. A magnetic field of 7.0 G applied along the propagation direction of the R1 Raman beam defines the quantization axis and 
destabilizes the dark states in the $^{2}S_{1/2}$ and $^{2}D_{3/2}$ manifolds for efficient Doppler cooling. 
We use the standard optical pumping and resonance
fluorescence state detection technique~\cite{2007PRA} to initialize and detect the ion's quantum state.
The ``dark" and ``bright" states used in state detection are 
$|0\rangle \equiv |S_{1/2}, F=0, m_{F}=0\rangle$ 
and $|1\rangle \equiv |S_{1/2}, F=1, m_{F}=0\rangle$, respectively. 
One of the ions is pumped into  
a metastable $^{2}F_{7/2}$ state with two photons, 
one from the  369~nm Doppler cooling laser and the other at 386.8~nm 
from a femtosecond modelocked pulsed laser that drives the 
$^2 P_{1/2} \rightarrow (7/2, 2)_{3/2}$ transition.

The ponderomotive trapping potential is generated by applying 
a rf signal at 30 MHz to two diametrically opposite rods of the trap 
(Fig.~\ref{fig:setup}b).
The radial trapping frequency is actively stabilized using a signal from a pickup 
coil that is positioned outside the vacuum chamber at a distance 
of around 5~cm from the trap. The systematic drift of the radial trapping frequency
$\omega_x/2\pi$ is less than $200~\mathrm{Hz/hour}$. 
The axial trapping frequency is independently controlled by DC voltages applied to 
the end caps and has negligible systematic drift. 

We also measure the coherence time of the phonons in the axial and radial out-of-phase 
modes with a  Ramsey experiment~\cite{Roos_shift_2008} to be 55(7)~ms and 10.2(9)~ms, respectively. The latter is limited by trapping frequency instability and can be 
extended to 38(4)~ms in spin-echo type measurements. 

\subsection{Raman beams}
The Raman beams are produced by a frequency-doubled mode-locked Ti:Sapphire laser (pulse 
duration 3~ps, repetition rate 76~MHz). The resulting beam has a central
wavelength of 374~nm and an average power of 250~mW. 
It is then split into three, sent through three acousto-optical modulators (AOMs), and
focused to beam waists of around $15~\mu\mathrm{m}$ at the ions' position as shown in
Fig.~\ref{fig:setup}{a}.
The R1 and R2 beams form 45$^\circ$ and 135$^\circ$ angles,
respectively, with the $z$-axis, while R3 is counter-propagating to R1.
The path lengths of all three beams are matched 
to a precision much better than the picosecond pulse length. 
The polarizations of the R1 and R2 beams are linear and mutually orthogonal, 
and the R3 beam polarization is parallel to that of R1.

In this configuration, the R1-R2 (R2-R3) pair couples the axial (radial) motional mode to
the internal state of the ion by driving stimulated Raman 
transitions~\cite{resolvedsbc,Hayes_2010}. 
By adjusting the Raman detuning, 
we may drive the ``carrier" ($|0\rangle|n\rangle \rightarrow |1\rangle|n\rangle$), 
the ``red" ($|0\rangle|n\rangle \rightarrow |1\rangle|n-1\rangle$) 
or ``blue" ($|0\rangle|n\rangle \rightarrow |1\rangle|n+1\rangle$) sideband transitions.
Here the first ket state corresponds to ion internal state, 
and the second to the state of its motional mode.

Moreover, when the R1 and R2 (R2 and R3) beams are detuned from each other, 
the polarization gradient of the resulting optical lattice at the location of 
the ion oscillates. This leads to a periodic optical dipole force applied to the ion 
in the state $|2\rangle \equiv |S_{1/2}, F=1, m_{F}=1\rangle$. The ion in the state $|0\rangle$ does 
not experience the force because its Stark shift is independent of
polarization~\cite{microwave}. 

\subsection{Control of normal mode frequencies}
The axial trapping frequency is controlled by DC voltages applied to the end caps,
and the radial trapping potential is generated by 30~MHz rf signal connected to the
diametrically opposite rods in the $x$ direction.
The radial trapping frequencies are shifted by DC offset voltages on the level of
hundreds of millivolts applied to the $x$ electrodes such that only the radial mode along
the $x$ direction interacts with the axial mode, while the radial mode along the $y$ direction is
far off resonance ($\lvert\delta\rvert/2\pi>200~\mathrm{kHz}$). 
The radial trapping frequency $\omega_r$ can be slowly changed between two values with
the help of two identical low-pass RC filters (LPF) that have a time constant of 2~ms, 
or rapidly changed with a time constant of 20~$\mu$s using different pairs of filters. 

The fast time scale (20~$\mu$s) is much shorter than the inverse coupling strength
$2\pi/\xi$, while the slow time scale (2~ms) is much longer than $2\pi/\xi$ to satisfy 
the adiabaticity criterion. Both timescales are much larger than one oscillation period
of the ion crystal, and we have experimentally verified that no significant 
motional excitations are induced during these frequency  sweeps~\cite{2012adiabaticsbcDrewsen,2014adiabaticsbcJapan}.

\subsection{Phonon mapping efficiency}
In our setup, the mapping of the phonon to the internal state of the ion is achieved by
applying a $\pi$-pulse on the red sideband that simultaneously removes one phonon from
the motional mode and changes the internal state of the ion. This procedure is not perfect,
and limited by the residual population of the other motional modes, power stability of
the Raman lasers and the internal state detection efficiency. 
To satisfy the condition $W(\alpha) \rightarrow 0$
for large $\alpha$, we determine the phonon mapping efficiency to be 
$\eta = 0.86 \pm 0.01$. This measurement agrees well with the result of an alternative
method, where the radial mode is initialized as a single-phonon Fock state and the
probability of making a spin flip is subsequently found to be $\eta = 0.89 \pm
0.04$. 

\subsection{Preparation of quantum states}
To prepare the coherent states shown in Fig.~\ref{fig:wigner}{b,c}, 
we start with all the motional modes of the ion crystal cooled to 
the ground state (Fig.~\ref{fig:wigner}{a}) and the $S_{1/2}$ 
ion pumped to the state $|0\rangle$. 
A microwave $\pi$-pulse drives the transition $|0\rangle \rightarrow |2\rangle$.
The optical lattice, formed by the R2 and R3 beams and running at the frequency
$\omega_r$, applies a force to coherently excite the ion motion~\cite{microwave}. 

The displacement in phase space $\alpha$ is calibrated by measuring
the average number of phonons $\bar{n}$ in the radial mode as a function of the coherent excitation duration. The measurement yields $|\alpha|^2=\bar{n}=3.0(2) * 10^{-4}*t^2$, 
where $t$ is expressed in microseconds.

To prepare the Schr\"{o}dinger cat states shown in Fig.\ref{fig:wigner}{d-f}, we follow
the method similar to~\cite{Monroe_cat_1996}. We prepare the $S_{1/2}$ ion in the
state $(|0\rangle + |2\rangle)|0\rangle_r / \sqrt{2}$ 
by the microwave $\pi / 2$ pulse and then apply a 
spin-dependent optical dipole force on resonance with the 
radial mode of motion that only displaces the ion in the internal state 
$|2\rangle$~\cite{microwave}. After that, we swap the internal states of the ions 
with a microwave $\pi$-pulse and apply the force with the phase shift $\phi$ to produce 
the $(|0\rangle |\alpha \rangle_r + |2\rangle | \alpha e^{i \phi}\rangle_r ) / \sqrt{2}$ 
state. Finally, we apply a microwave
$\pm\pi / 2$ - pulse to the internal state of the ion and arrive at the state 
$ |0\rangle (|\alpha \rangle_r \pm  |\alpha e^{i\phi} \rangle_r)/2 
+ |2\rangle ( |\alpha \rangle_r \mp | \alpha e^{i\phi}\rangle_r ) / 2$.
We then detect the internal state of the ion using the standard fluorescence technique. 
If the ion is found in the $|0\rangle$ internal state, the ion has scattered 
no photons and the motional state of the ions is projected onto the Schr\"odinger cat state $(|\alpha \rangle_r \pm |\alpha e^{i\phi}\rangle_r)/\sqrt{2}$. 
If the ion is found in the ``bright" state $|2\rangle$, the motional state is 
destroyed by the photon recoil and we omit these cases.

The Fock states shown in Fig.~\ref{fig:wigner}{g-j} are generated by the following sequence: a $\pi$-pulse on the blue sideband of the $|0\rangle \rightarrow |1\rangle$ transition that
adds a phonon to the motional mode, followed, if necessary, by another $\pi$-pulse on
the red sideband  that adds another phonon, or a $\pi$ pulse on the carrier transition that
returns the ion to the initial internal state. To generate the $n$-phonon Fock state
$|n\rangle_r$, $n$ sideband pulses are applied.

\end{document}